# Fermentation based carbon nanotube bionic functional composites

Luca Valentini,[1]* Silvia Bittolo Bon,[1] Stefano Signetti,[2] Manoj Tripathi,[3] Erica Iacob,[3] Nicola M. Pugno[2,3,4]*

*1 Dipartimento di Ingegneria Civile e Ambientale, Università di Perugia, UdR INSTM, Strada di Pentima 4, 05100 Terni - Italy.*

*Tel: +39 0744 492924; E-mail: luca.valentini@unipg.it*

*2 Laboratory of Bio-Inspired and Graphene Nanomechanics, Department of Civil, Environmental and Mechanical Engineering, University of Trento, Trento - Italy*

*3 Centre for Materials and Microsystems, Fondazione Bruno Kessler, Trento - Italy*

*4 School of Engineering and Materials Science, Queen Mary University of London, Mile End Road, London - United Kingdom.*

*Tel: +39 0461 282525; E-mail: nicola.pugno@unitn.it*

**Abstract**

The exploitation of the processes used by microorganisms to digest nutrients for their growth can be a viable method for the formation of a wide range of so called biogenic materials that have unique mechanical and physical properties that are not produced by abiotic processes. Based on grape must and bread fermentation, a bionic composite made of carbon nanotubes (CNTs) and single-cell fungi, the *Saccharomyces cerevisiae* yeast extract, was prepared by fermentation of such microorganisms at room temperature. Tensile tests on dried composite films indicated a strong increment of the mechanical properties with respect to the neat yeast film, in terms of strength, Young's and toughness moduli. The mechanical results have been rationalized in terms of a CNT cell bridging mechanism where the strength of the composite is governed by the adhesion energy between the bridging carbon nanotubes and the matrix. The addition of CNTs also significantly improved the electrical conductivity of the fermented yeast film along with a higher photoconductive activity when it was exposed to solar light. We demonstrate that this simple cellular factory can be used to produce bionic functional composites in eco-friendly, facile and bionic efficient way.

**Introduction**

Microorganisms play an essential role in the biogenic of elements and in the formation of materials with unexplored properties.[1-3] Biogenic materials are formed in the nanometer scale through diverse metabolic activities and by passive surface reactions on cell walls or extracellular structures. For example, it is known that acetobacter bacteria spins cellulose – a byproduct – as it consumes glucose, the reasons for which are unclear but it is thought the material might protect the bacteria colony from external contamination.[4] Extracellular polymeric secretions are multipurpose polymers that are important for applications in several fields. Extracellular polymeric secretions are involved in cellular associations, bacterial nutrition, and interaction of bacteria with their bio-physicochemical environment. Thus to create the material, the microorganism requires specific conditions.

Yeast is a cellular factory which is able of taking simple molecules from its environment, such as sugars, and synthesize new elements needed for its growth at a mild temperature.[5-8] Moreover, many of such microorganisms during their growth have the ability to adhere and to form a biofilm on different kinds of surfaces in nature.[8,9] It would be expected that a solution of nanomaterials and water, for example, can be mixed with nutrients for the microorganisms and also gives novel outstanding properties to the end material after the digestion process has taken place.[10-12] Fermentation of microorganism in the presence of nanomaterials could be a viable method to create a biogenic composite. It could be possible to get from large cultures enough material to create composites with unexpected properties.[13]

Yeast cells are widely used in industrial and biological processes. The beer´s yeast, *Saccharomyces cerevisiae*, is the yeast responsible for sugar fermentation and has been used for centuries in wine and bread making. Moreover beer yeast extract is inexpensive, non-toxic, easy to prepare and abundantly available in nature.[14,15] In fermentation, flocculation commonly occurs when the sources of fermentable sugars are exhausted.[16] During the flocculation the yeast cells aggregate rapidly in the medium in which they are suspended forming a film after the removal of the liquid medium. Flocculation of microorganisms is simple, cheap and affordable for thin film deposition because it does not require complex procedures. However, to the best of our knowledge, the spontaneous incorporation of nanomaterials in the inner structure of yeast cells during the fermentation, has not been achieved to date.

Carbon nanotubes (CNTs) are nowadays materials produced on industrial scale,[17] making them a tunable platform for mimicking what occurs in nature when cells grow on nanostructured surfaces. The mechanism of nutrition and growth of microorganism represents an unexplored field to design interface between microorganism and such nanomaterials. In several recent studies, CNTs were shown to be a very useful electron transfer for enzymatic as well as cellular applications.[18] Considering the conducting nature of CNTs, the interactions between microorganism and carbon nanotubes could be used for generating bionic nanomaterials with unexpected electrical properties. Several cells (neurons, muscle) were found to respond to electrical signals but unfortunately are difficult to culture with procedures that are time consuming.[19,20] *Saccharomyces cerevisiae* is instead more robust and accessible fungi cell that allow fast and reliable experiments to investigate their electrical behavior when combined with conducting nanomaterials. Hence, we decided to explore their possible use in electrical applications and particularly to examine whether yeast after its fermentation and flocculation can interact in synergy with CNTs still benefiting from their electrical conductivity, high surface area, flexibility and ability to be wrapped by microorganisms.

Here, we report the production of beer's yeast cells/carbon nanotube composite directly by fermentation of the yeast extract in presence of CNT aqueous dispersion. We observe in the composite an increment of the mechanical properties with respect to the neat fermented yeast, in terms of tensile strength, Young's and toughness moduli. The electrical and optical analysis demonstrated that fermentation of the beer's yeast in presence of CNTs enhances the electrical conductivity as well as the photoconducting activity of the composite film. Such results could open the way to the realization and exploitation of such bionic structures for the realization of novel functional materials.

**Experimental Details**

Carbon nanotubes (NC 7000) were purchased from Nanocyl and their structure was confirmed by transmission electron microscopy (see Supporting Information) (average diameter 9.5nm, average length 1.5µm). *Saccharomyces cerevisiae* based commercial beer yeast extract with additives was used as the medium for fermentation. Water solution (50 mg/ml) of yeast was prepared in a sterilized flask at 110 rpm and 30°C for 1h. After that, sugar (i.e. sucrose) was added for the fermentation. The amount of sugar added is usually between 3 and 5 times the weight of the medium. Neat and fermented yeast were then drop into sterilized circular Al molds and the liquid medium was left to evaporate at 30°C during the night. Films of circular shape were then obtained.

Water dispersion of CNTs (1mg/ml) were prepared by tip sonication. The dispersion of CNTs (1 mg/ml) was then added to the yeast solution and stirred at 110 rpm at 30°C for 1 h. Then the yeast/CNTs solution was put into a sterilized circular Al mold, and the liquid medium was left to evaporate at 30°C during the night. In another sterilized flask the same procedure was adopted by adding sugar to promote the fermentation of the yeast/CNTs solution (Figure 1a). After fermentation, the liquid medium was dried in a sterilized mold and a composite film was obtained.

Field emission scanning microscopy (FESEM) was used to investigate the cross section of the samples obtained by fracture in liquid nitrogen. Morphological characterization was carried out on films deposited on Si substrates by means of an AFM apparatus (P47H solver from NT-MDT) in intermittent contact mode at room temperature in air condition. Silicon cantilever was used of approximately tip radii 25nm covered with native oxide.

The tensile properties of films, i.e. tensile strength (TS), fracture strength (FS) and elongation at break (EB), were measured using a universal tensile testing machine (Lloyd Instr. LR30K) with a 50 N static load cell. The film samples were cut into strips (30 mm × 12 mm). The gauge length was 20 mm, and the extension rate was set at 2 mm/min.

Ultraviolet–visible (UV–Vis) measurements of the deposited films were carried out with a Perkin-Elmer spectrometer Lambda 35; for all samples, a neat quartz slide was used as the reference. The optical absorbance was obtained on films of the same thickness (~5 µm).

The current–voltage characteristic was performed by a computer controlled Keithley 4200 Source Measure Unit. The electrical conductivity of the samples was monitored, at room temperature, by applying a sweeping DC electric voltage from -40 V to 40 V between the electrodes. Forward and reverse scans were performed on each sample. The photocurrent was measured under AM1.5D 150 mW*cm$^{-2}$ illumination from a Thermal Oriel solar simulator. Photoelectrical measurements were obtained for the films over several on/off light illumination cycles.

**Results and Discussion**

The scheme of our method is sketched in Figure 1; the yeast cells gain nutrient from the conversion of sucrose and grow with the division process in presence of sucrose/CNTs aqueous dispersion. Figure 1 shows the appearance of the prepared samples. It can be seen that the fermented yeast presents a yellowish brown color similar to that of beer yeast. The composite films exhibit a black color. Based on visual observation, the incorporation of CNTs led to the formation of a brittle film.

The effect of fermentation and CNT addition on the morphology of the films was investigated by AFM and the results are reported in Figure 2. The shape of the cells was found to be spheroidal. In general, the fermentation process of the yeast decreases the average surface roughness from 0.3 μm to 0.2 μm (Figures 2a) and 2b)) accordingly to the yeast morphology reported elsewhere.[21] Representative images of the samples (see Figure 1S) show a morphology consisting of aggregated yeast cells with cell boundaries being the surface of the film obtained by fermentation of yeast with CNTs more homogeneous (i. e. lower protrusions). Under FESEM with an acceleration voltage of the electron beam set to 5 kV, it can be seen clearly that almost all the yeast cells after fermentation kept their spherical shape structures (Figure 2S) with a morphology comparable with those obtained by AFM. It can be seen also from FESEM analysis that yeast cells before fermentation appeared concave and burst. This can be attributed to hyper-osmotic shock by exposure in the high vacuum and under the electron beam of FESEM.

The strategy adopted for this study is based on the previous papers reporting that *Saccharomyces cerevisiae* multiplies with a process where a daughter cell is initiated as growth from the mother cell.[22] The round protrusion on the cell surface, visible in Figure 3a, is a bud scar. The bud scar forms on the cell after the process of division has taken place. We exploit this growth process in order to verify if intracellular transport of carbon nanotubes takes place during this growth process as sketched in Figure 1b. Stressing the composite sample by prolonged exposure under high vacuum condition and electron beam of FESEM, suggests that in the composite film obtained after fermentation the CNTs are bridging the yeast cells and are internalized by them (Figures 3b and 3c). Such results are in agreement with those found by Zhou et al.[23] who showed that CNTs did not remain on the outside membrane but transverse the membranes and localized in the cell.

Microscopy studies demonstrate that this approach can be regarded as the effective and versatile way to create hybrid micro-bubbles. After the FESEM inspection, we suggest that the bridging promoted by the internalized CNTs might result in additional mechanical properties to those composites accordingly to a pull-out sketch reported in Figure 4. In this regard tensile tests were performed and the stress-strain curves of the prepared samples are reported in Figures 5a and 5b. The values of fracture strength (i. e. the stress at the ultimate strain), the elongation at break (i. e. the ultimate strain), the toughness (i.e. the area underlying the stress-strain curves) and elastic moduli are reported in Figure 3S. Generally, it was observed that the fracture strength, toughness and elastic modulus increase when CNTs were added to the yeast. The incorporation of CNTs before fermentation demonstrated to produce the highest variation in tensile properties of the composite film, being this effect highest for fracture strength; this led to extremely fragile film with the lowest elongation break value (see Figure 5b and Figure 3S). This result is consistent with our finding reported above based on visual appearance of the composite film on the mold as reported in Figure 1. On the contrary, the composite film obtained after fermentation showed both a lower fracture strength and elastic modulus with a significantly higher elongation at break (Figure 5b and Figure 3S). The increase of elongation at break after fermentation, can be attributed to a decrease in film density and thus to an increase of the mean free volume of the film.[24,25]

The tensile test results can be rationalized assuming a pull out model[26-28] representing the failure mechanisms of bridging nanotubes as sketched in Figure 4. The tensile strength of the composite, $\sigma_f$, can be expressed as the combination of the strength of the matrix, $\sigma_{f\text{-yeast}}$, and the pull out strength of CNTs, $\sigma_{f\text{-cntpo}}$, accordingly to the following mixture equation:

$$\sigma_f = f\sigma_{f\text{-cntpo}} + (1-f)\sigma_{f\text{-yeast}} \tag{Eq.1}$$

where $f$ is the volume fraction of CNTs bridging the interface of the yeast cells (see Figure 4). For $f = 0$ we get the strength of the matrix (i.e. $\sigma_f = \sigma_{f\text{-yeast}}$). We assume from FESEM observation a $f$ value of

0.0205 and tensile strength values of 0.240MPa and $0.297*10^{-1}$ MPa for $\sigma_f$ (i. e. the fermented composite) and $\sigma_{f\text{-yeast}}$ (i.e. the fermented yeast), respectively, obtaining a $\sigma_{f\text{-cntpo}}$ of 10.29 MPa. From this calculation we found that the $\sigma_{f\text{-yeast}} \ll \sigma_{f\text{-cntpo}}$ indicating that the strength of the composite is governed by the adhesion energy between the bridging carbon nanotubes and the matrix.

Assuming that the applied stress is transferred to the nanotube via a nanotube–matrix interfacial shear mechanism at the molecular level, the pull out force of the CNTs, $F_{f\text{-cntpo}}$, may be expressed as it follows:

$$F_{f\text{-cntpo}} = 2\pi r * G'_{f\text{-cntpo}} \qquad (Eq.2)$$

where the adhesion energy, $G'_{f\text{-cntpo}}$, can be calculated from the pull out strength and the radius of the outer shell, r, of CNTs as

$$G'_{f\text{-cntpo}} = \sigma_{f\text{-cntpo}} * (r/2) \qquad (Eq.3)$$

We can thus obtain the adhesion energy $G'_{f\text{-cntpo}}$ from the pull out strength as described above considering the mean value for the nanotube radius of ~4.7 nm. The $G'_{f\text{-cntpo}}$ value obtained is thus ~0.024 N/m.

A simple system of two cells before and after fermentation was modelled via finite element simulation in order to understand the effect of the incorporation of CNTs within the yeast. The image of Figure 3b shows that the shape of the cells within the ensemble can be approximated as hexagonal prism. The dimension of the cell, mother and daughter, and their elastic properties as well, were taken from the work of Ahmad et al. [21]. They report for the mother cell a diameter $d_m = 5.565$ μm and Young's modulus $E_m = 1.46$ MPa, while for the daughter cell $d_d = 4.467$ μm and Young's modulus $E_d = 1.10$ MPa. The cell dimension measurement from this work are consistent with our estimations (see Figure 2 and Supplementary Figure 1S). The hexagonal base of the prism was dimensioned in order to have an equivalent area of a circle of diameter *d* while the cell height is finally univocally determined from the cell volume reported elsewhere.[21] The cells are modeled with under-integrated solid elements with spurious mode stabilization.[29] The load is applied as imposed displacements on the lateral face of the cells in order to get the post critical regime.

The interface between the two cells was modeled via a cohesive zone model (CZM) based contact [29] for which the overall adhesion energy, assumed of pure Mode I fracture, is a combination by a rule of mixture of the fracture energy of the pristine yeast interface and the adhesion energy of CNTs determined above with the pullout model. From Figure 5a the experimental curve slope and fracture strength (see also Supplementary Figure 3S) are lower than the cell elastic parameters reported in literature [21,30], thus we deduce that the failure occurs mainly due to interface opening mechanism. The curve was then used to determine the yeast interface fracture energy that is estimated to be $G_{f\text{-yeast}}$ = 0.0193 N/m assuming a cell diameter, and then distance from interfaces, of 5 µm. This value has to be compared with the pull out energy of CNTs projected on the contact interface:

$$G_{f\text{-cntpo}} = G'_{f\text{-cntpo}} \cdot (l/r) \tag{Eq. 4}$$

estimated to be 7.72 N/m where *l* is nanotube length, thus 1.5 µm, assuming that on average the dissipation occurs over a length *l*/2 since not all nanotubes present an ideal anchorage *l*/2-*l*/2 within the two cell membranes. Thus the total adhesion energy of the composite interface is:

$$G_{f\text{-comp}} = f \cdot G_{f\text{-cntpo}} + (1-f) \cdot G_{f\text{-yeast}} \tag{Eq. 5}$$

from which is possible to estimate the CNTs volume fraction *f* by minimizing the difference between the energy release between the simulation derived and the experimental curves. The constitutive contact laws $\sigma_i$ -$\delta_i$ for the two phases (generically *i*) are assumed to be bi-linear:

$$\sigma_i = \sigma_{f\text{-}i} \cdot (\delta_i/\delta_{f\text{-}i}) \qquad \text{for } \sigma_i < \sigma_{f\text{-}i} \qquad (Eq.\ 6.a)$$

$$\sigma_i = \sigma_{f\text{-}i} \cdot (\delta_{u\text{-}i} - \delta_i)/(\delta_{u\text{-}i} - \delta_{f\text{-}i}) \qquad \text{for } \sigma_i > \sigma_{f\text{-}i} \qquad (Eq.\ 6.b)$$

where $\delta_{f\text{-}i}$ is the crack opening corresponding to $\sigma_i = \sigma_{f\text{-}i}$ and $\delta_{u\text{-}i}$ is the ultimate crack opening at which $\sigma_i = 0$. In particular for the CNTs pull-out is $\delta_{f\text{-}cntpo} = l/4$ and $\delta_{u\text{-}cntpo} = l/2$ according to the pull-out model presented above, and $\delta_{f\text{-}yeast} = 0.54$ µm and $\delta_{u\text{-}yeast} = 0.88$ µm for the yeast interface estimated from the experimental curve of Figure 5.a. For the composite interfaces the rule of mixture is applied to both characteristic stresses and crack openings above defined, analogously to Eq.1 and Eq.5. Finally the failure of the interaction at each contact nodal point occurs when the current energy release, G, overcomes the critical value:

$$(G/G_{f\text{-}comp}) > 1 \qquad (Eq.\ 7)$$

Figure 5a and 5b shows the superposition of our simulation on experimental curves. In particular Figure 5a shows the results for mother-daughter cells in fermented yeast, while in Figure 5b the cases of equal cells with CNTs before fermentation and mother-daughter cells with CNTs after fermentation were considered. The obtained results are in good agreement with experiments, following the variation in the peak tensile stress and ultimate strain. In particular the best match between the simulated and real curves occurs for volume fraction values of $f_{y+CNTs} = 0.071$ and $f_{y+CNTs,ferm} = 0.019$ for yeast/CNTs and fermented yeast/CNTs composites, respectively. Note that the latter value is comparable with the one estimated from the SEM images (Figure 4) and assumed in the pull-out model. These findings may indicate that the increase of the cell volume after fermentation results in a higher contact area of the yeast interface or in a different CNTs pull-out length, dropping the volume fraction of the nanotubes and thus explaining the lower failure stress and the higher failure strain observed for the composite film after fermentation. This suggests further investigations for understanding the interface transformation mechanism upon fermentation.

Once collected the morphology and the mechanical properties of the films, we next investigated the electrical characteristics of the composite films, and assess how they compare with the native films made of neat yeast. This filled and bridged microenvironment could be a viable network for enhancing the electron transfer through the nanocomposite.

Thus we investigate whether there is any advantage to add CNTs to the yeast in terms of electrical conductivity. Current-voltage characteristic was thus used to characterize the conductivity of yeast modified with CNTs as reported in Figure 6. We observed that the increase in conductivity is more prominent with the CNTs. We further observed that the addition of sucrose to the neat yeast resulted in a better conductivity being this effect more pronounced for the fermented yeast/CNTs system. In attempting to define the source of this behavior, we hypothesized that the bridging of CNTs of the yeast during the fermentation improved the percolation pattern for electron transfer to the electrodes, via the conducting capabilities of CNTs. The peaks observed for forward and reverse voltage scans for the fermented samples, could be attributable to the reduction and oxidation activity of the sucrose oxidase in the composite films.[31] Further investigations into electrochemical properties of such systems are underway.

Light absorbance in the UV–Vis range of the prepared samples are presented in Figure 7a. In the visible range (350–800 nm), depending on both the CNTs incorporation and fermentation process, differences were obtained in terms of normalized optical absorbance. The absorbance value for the yeast film incorporating CNTs relative to neat yeast shows a significant increase. This difference may be associated with the presence of an extended network of aggregated CNTs that blocked passage of visible light in a more effective way. The fermentation of the yeast in presence of CNTs resulted in a decrease of the optical absorbance indicating a better dispersion and/or confinement of the CNT bundle, thus resulting in a better transparency of the composite film.

In previous studies food pigments extracted from yeast fermentations have been studied as a novel sensitizing dye for dye-sensitized solar cells.[32-34] Moreover, Hildebrandt et al.[35] recorded photocurrents and photovoltages with yeast plasma membrane attached to a planar lipid membrane and to a polytetrafluoroethylene (Teflon) film, respectively. In Figure 7b the photocurrents that evolve before and after light exposure on the prepared samples are reported. It is clear from the curves that neat yeast as well as fermented yeast showed a current increase under illumination. The spectral irradiance for a Xenon lamp between 400 and 500 nm under the AM1.5D standard condition is about 16.9%; the absorption spectrum of the films in this optical window (i. e. 416nm from Figure 7a) supports the observation that a photocurrent signal for the yeast extract has been recorded.

It is important to note that when the same experiment was conducted in the presence of CNTs, the baseline of the dark current signal increases being this effect more pronounced for the composite film obtained after fermentation. The photo response, estimated as the relative variation between the dark and the light signals, increases from 16% to 26% when the fermented yeast was compared with the composite obtained after fermentation. A possible mechanism could be related to a lower surface roughness of the composite film obtained by fermentation that promotes a more favorable charge transfer from the yeast cell to the CNTs.

**Conclusions**

In this work composite films were successfully prepared by the fermentation of beer yeast in presence of carbon nanotubes. The fermentation of beer yeast in presence of CNTs produced marked changes in mechanical properties of the composite film. The composite resulting from fermentation showed an improved electrical conductivity along with an enhanced photoconducting activity. In summary we demonstrated the development of novel structures by means of biogenesis without the post-sythesis addition of the inorganic components by physical or chemical methods. In this context the present work aims the development of future implication of such bionic materials in the manufacture of physical smart objects with multifunctional properties in the fields of functional tissues, energy storage and light harvesting applications.

**Acknowledgments**

NMP is supported by the European Research Council (ERC StG Ideas 2011) BIHSNAM n. 279985 on "Bio-Inspired hierarchical supernanomaterials", ERC PoC 2013 KNOTOUGH n. 632277 on "Supertough knotted fibers", ERC PoC 2015 SILKENE nr. 693670 on "Bionic silk with graphene or other nanomaterials spun by silkworms"), by the European Commission under the Graphene Flagship (WP10 "Nanocomposites", n. 604391) and by the Provincia Autonoma di Trento ("Graphene Nanocomposites" , n. S116/2012-242637 and delib. reg. n. 2266). SS acknowledges support from BIHSNAM.

**Author Contributions**

L.V. had the idea, designed and supervised the entire research and analyzed the results. S.B.B. prepared the samples and performed the morphological, electrical and UV-Vis characterizations. M.T. and E.I. performed AFM analysis and scratch tests. S.S. performed FEM simulations. S.S. and N.M.P. rationalized the mechanical results. All authors have revised and given their approval to the final version of the manuscript.

**Additional Information**

The authors have no competing interests as defined by Nature Publishing Group, or other interests that might be perceived to influence the results and/or discussion reported in this paper.

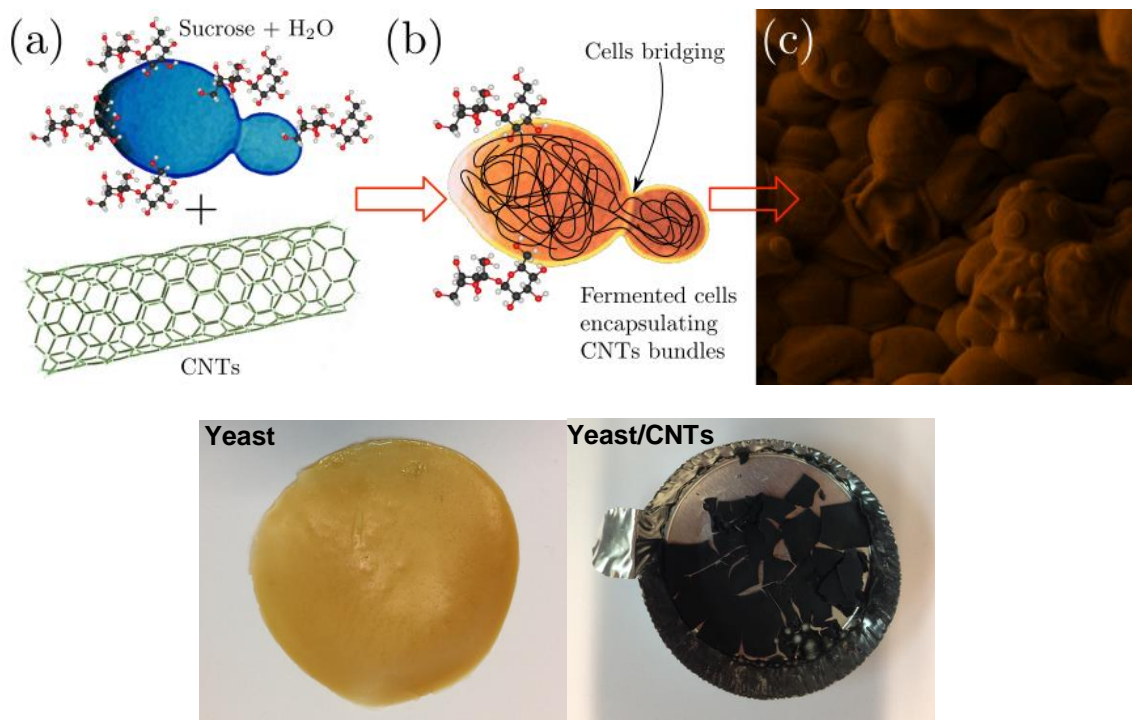

FIG. 1. Schematic representation and appearance of the fermentation-mediated assembly of carbon nanotubes and yeast cells.

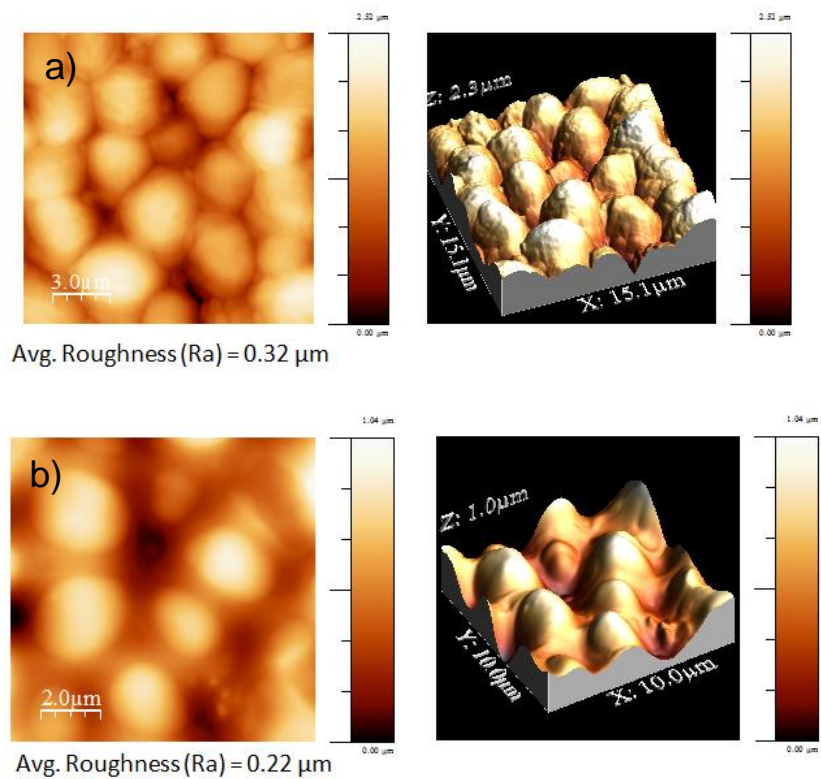

FIG. 2. 2D and 3D topography AFM images and average surface roughness ($R_a$) of (a) yeast/CNTs and (b) fermented yeast/CNTs composite films.

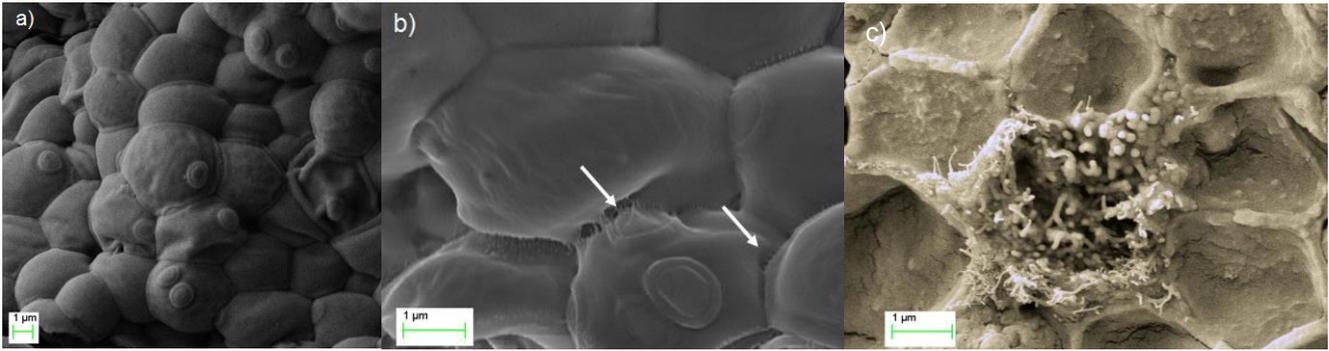

FIG. 3. (a) FESEM image of *Saccharomyces cerevisiae*. (b) FESEM image showing CNTs bridging yeast cells. The arrows indicate the CNTs bridging the yeast cells. (c) FESEM image of the cross section of the fermented yeast/CNTs film after prolonged exposure to FESEM where CNTs protruding from a broken yeast cell are visible.

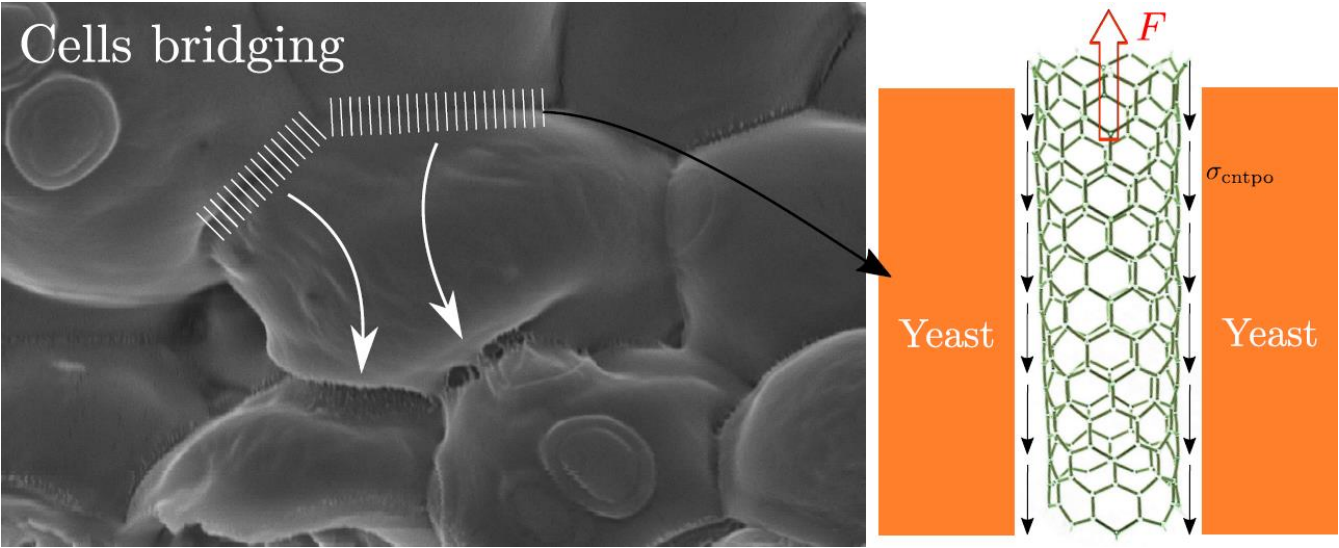

FIG. 4. Schematic view of the bridging effect on a SEM image of the fermented yeast considered for the pull out model and finite element simulation.

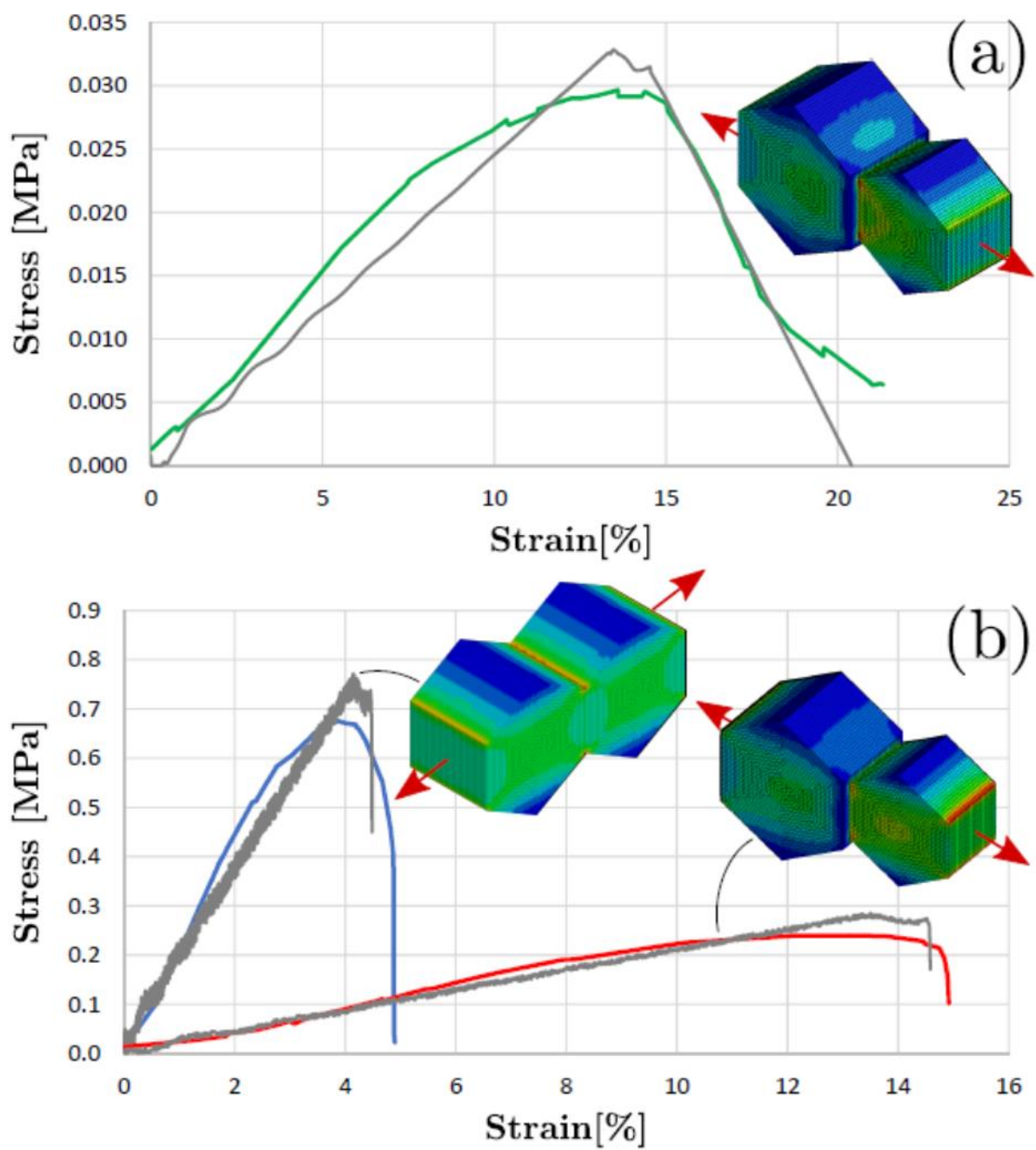

FIG. 5. Stress-strain curves obtained from tensile tests on (a) fermented yeast sample (green curve) and (b) yeast/CNT composites prepared before (blue curve) and after (red curve) fermentation, respectively. The comparison with the result of FEM simulation of strain of a two-cell system are also reported.

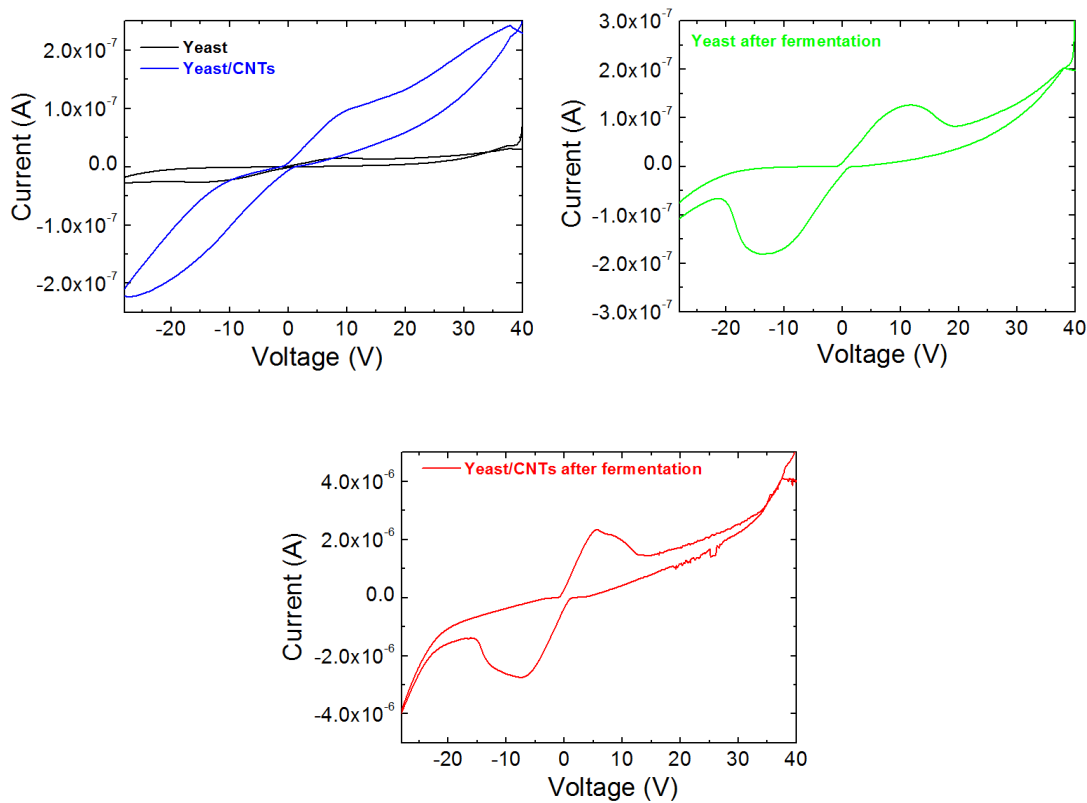

FIG. 6. I-V characteristics of the prepared samples.

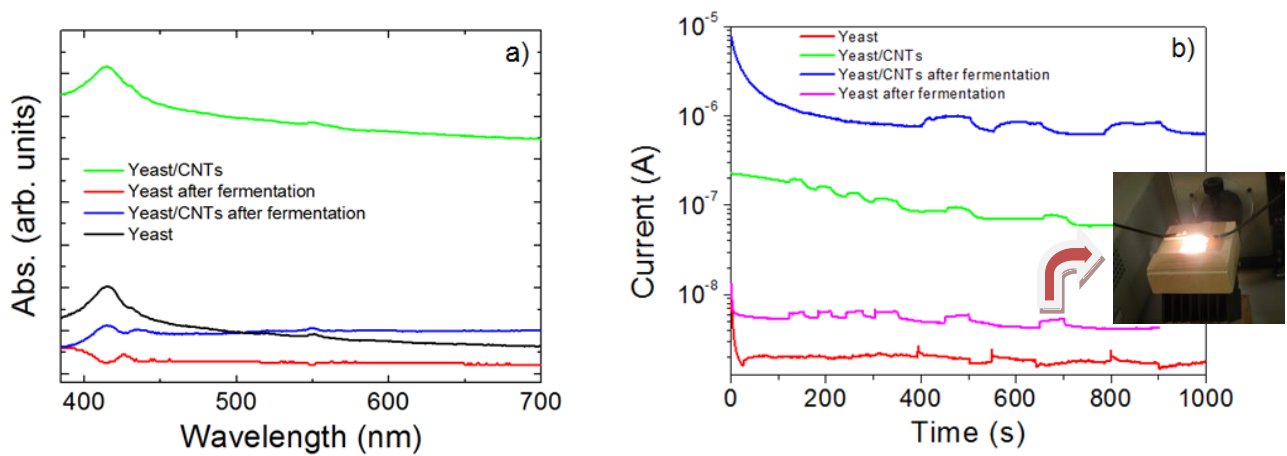

FIG. 7. (a) UV-VIS spectra of the prepared films. (b) Photocurrent recorded on the prepared samples when the films were exposed to illumination cycles. The change of the current intensity represents the switching on and off of the solar simulator (i. e. 3, 6, 6 and 3 cycles for the yeast, yeast after fermentation, yeast/CNTs and yeast/CNTs after fermentation, respectively).

23